# Comment on "Enhanced transmission through periodic arrays of subwavelength holes: the role of localized waveguide resonances"

Cheng-ping Huang and Yong-yuan Zhu*

National Laboratory of Solid State Microstructures, Nanjing University

Nanjing 210093, P.R. China

Recently, Ruan et al. [1] numerically studied the enhanced light transmission through metal films perforated with rectangular holes, where the hole size was fixed and lattice constant was varied instead. Besides a flat band structure, it has been suggested that the longer transmission peak is corresponding to the cutoff wavelength of waveguide, and that it is mainly determined by the hole size and almost independent on the periodicity. Thus the transmission enhancement at the longer wavelength has been attributed simply to the localized waveguide resonance. Nonetheless, as shown in the following, the cases considered by authors are not general, thus the conclusion expanded is not correct.

For the perfect metal film with rectangular holes, it is well known that the cutoff wavelength of waveguide equals the twice of the longer hole side. Then when the period is large enough, the peak position which is longer than the period can be far from the cutoff wavelength [2]. For the hole size 300nm*400nm and period 1000nm, for example, the peak is over 1000nm whereas the cutoff is only 800nm. Thus, although the transmission peak of a single hole is related to the cutoff wavelength [3], the presence of periodicity will change this character greatly. Moreover, for the real metal film, the cutoff wavelength has not been calculated and compared with the transmission peak [1]. Recently, we have developed an analytical model for the transmission, in which the propagation constant is expressed as [4]

$$q_0 = \sqrt{k_h^2 - \alpha^2 - \beta^2}. \qquad (1)$$

Where $k_h = k_0\sqrt{\varepsilon_h}$ ($\varepsilon_h = 1$ for air), $\alpha$ and $\beta$ are determined by

$$\tan\frac{\alpha a}{2} = \frac{k_0 \varepsilon_h}{i\alpha\sqrt{\varepsilon_m}}, \quad \tan\frac{\beta b}{2} = \frac{k_0\sqrt{\varepsilon_m}}{i\beta}. \qquad (2)$$

Here $a$, $b$ is the length of rectangular hole side (the magnetic field is along the side $b$), and $\varepsilon_m$ is permittivity of the metal. With the equations (1) and (2), it is found that the cutoff wavelength of waveguide (75nm*225nm) is 726nm, still far from the peak position around 836nm (see Fig.4 of Ref. [1]).

On the other hand, for the fixed hole size, the longer transmission peak can be dependent obviously on the period. Actually, this point has been suggested in their Fig.4 (note that the wavelength axis is not homogenous [1]). To see this more clearly, we have calculated the transmission spectra with the analytical model [4] and the results are shown in Fig.1a. It can be seen that our results agree well with the numerical calculations. When the period is varied from 425 to 450 and 475nm, the shorter transmission peak locates respectively at 618, 655 and 698nm (their numerical results are 617, 653, and 694nm [1]), and the longer peak is shifted from 843 to 853 and 863nm. Further calculations show that the peak shift is sensitive to the aspect ratio of the hole. Fig.1b presents the transmission spectra of another rectangular holes (250nm*300nm), where a variation of period from 600 to 700nm gives rise to 105nm redshift of the longer transmission peak, which is much larger than 60nm of the shorter one (similar results have been observed experimentally by Degiron et al. [5]). This behavior holds even for the perfect metals. In this case, i.e., the aspect ratio is smaller and both transmission peaks vary strongly with the period (Fig.1b), the electromagnetic band structures associated with the two peaks will be similar to each other and a flat band structure will not be present. The results suggest that the longer transmission peak benefits usually from both localized waveguide mode and extended surface mode, thus excluding the so-called waveguide resonance.

This work was supported by the State Key Program for Basic Research of China (Grant No. 2004CB619003), by the National Natural Science Foundation of China (Grant No. 10523001 and 10474042).          * E-mail: yyzhu@nju.edu.cn

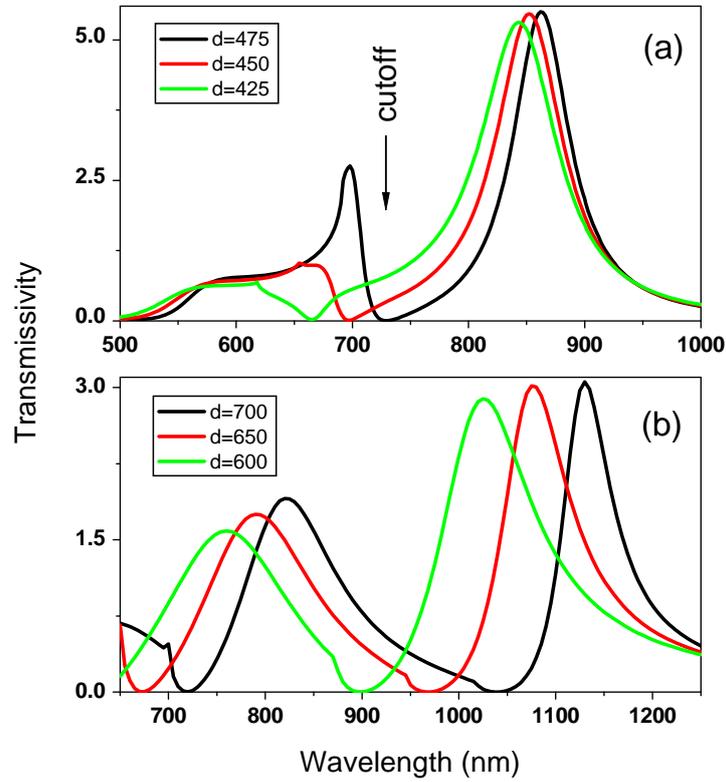

**Fig. 1:**

Zero-order transmission spectra of rectangular holes: (a) the hole size is fixed at 75nm*225nm, and the period is varied from 425 to 450 and 475nm (the arrow indicates the cutoff wavelength); (b) the hole size is fixed at 250nm*300nm, and the period is respectively 600, 650 and 700nm. Here the film thickness is 200nm, and the permittivity of glass substrate is 2.117.